\documentclass[twocolumn,showpacs,preprintnumbers,amsmath,amssymb]{revtex4}
\usepackage{amsmath}
\usepackage{graphicx}
\usepackage{dcolumn}
\usepackage{bm}
\usepackage{textcomp}

\newcommand{\w}{\omega}

\begin{document}

\preprint{APS/123-QED}

\title{Coherence-based multivariate analysis of high frequency stock market values}

\author{Donatello Materassi}
	\email{materassi@dsi.unifi.it; mater013@umn.edu}
  \altaffiliation[also at ]{University of Minnesota}
\author{Giacomo Innocenti}%
 \email{ginnocen@dsi.unifi.it}
\affiliation{Dip. di Sistemi e Informatica, Universit\`a degli Studi di Firenze\\via S. Marta 3, 50139, Firenze, Italy
}
\begin{abstract}
The paper tackles the problem of deriving a topological structure among stock prices from high frequency historical values.
Similar studies using low frequency data have already provided valuable insights.
However, in those cases data need to be collected for a longer period and then they have to be detrended.
An effective technique based on averaging a metric function on short subperiods of the observation horizon is suggested.
Since a standard correlation-based metric is not capable of catching dependencies
at different time instants, it is not expected to perform the best when dealing with
high frequency data.
Hence, the choice of a more suitable metric is discussed.
In particular, a coherence-based metric is proposed, for it is able to detect any possible linear relation between two times series, even at different time instants.
The averaging technique is employed to analyze a set of $100$ high volume stocks of the New York Stock Exchange, observed during March 2008.
\end{abstract}
\maketitle
\section{Introduction}
Deriving information from financial markets is a formidable and challenging task.
During the last decades many approaches have been developed and in recent
years physicists and engineers have proposed novel approaches \cite{mant95,ms96,introeconophysics,stan99b}.
It has already been proved that the time series of financial prices are related
and in this respect many correlation analyses have provided useful contributions
and insights \cite{tum07,nrm07}.
However, these analyses were carried on using daily (i.e. ``low frequency'') data,
while in the financial field a very important issue of the decision process
is to understand and to quantify the influences, which occur in a short time range.
Hence, procedures specifically designed to deal with ``high frequency'' data are
desirable.
Further motivations to prefer a short period analysis can be found in a reduced
prefiltering stage.
Indeed, economic processes are naturally organized into different time units
(days, weeks, months and years) and trends and seasonal factors
are expected to be less relevant at small time scales, i.e. when
the influence of short time financial phenomena is dominant.\\
In scientific literature and especially in the econometrics and
engineering fields several techniques have been developed to derive
structured information from the sampled data.
In particular, the black box approach is probably the most common paradigm
in the linear framework (see e.g. \cite{ksh00}).
On the other hand, in the ``high frequency'' scenario, spectral analysis seems
to provide a better approach to the problem, as it has recently been observed in \cite{pos08}.
In this paper we pursue a different approach, that has been recently introduced in  \cite{gid08} and that is based on Wiener filtering and frequency analysis.
In particular, such a method exploits linear modeling techniques to define dynamical laws describing in a qualitative and quantitative way the connections among the time series of the data set.
The main points of this approach will be introduced in Section~\ref{sec:theo fram}, while Section~\ref{sec:exp res} will be devoted to the application of the method to the stock price historical series, sampled at high frequency.
Finally, the results will be interpreted in terms of graph theory and 
the relative clusterization will be presented as well.
%
\section{Theoretical framework}
\label{sec:theo fram}
It is known that financial time series are non-stationary
and that, conversely, they exhibit behaviours affected by cyclical
and non-cyclical trends.
Thus, in most situations low frequency data
need to be detrended, that is removed of any deterministic
component in order to reconstruct them as stationary processes,
which can be studied in the standard statistical framework.
On the other hand, high frequency data, if observed over a short period,
can approximately be considered stationary, since the trend
components mainly present slow variations.\\
In this paper, we suggest a technique to exploit high frequency data
in the multivariate analysis described in \cite{gid08}.
In particular, we will introduce a strategy suitable
to deal with such high frequency data over long time horizons,
i.e. when the trend components can not be considered negligible.
Moreover, we want to underline that the method presented \cite{gid08}
turns out to be the ``dynamical'' extension of the procedure
originally described in \cite{mant95}.
In this respect, we find useful to recall that
in \cite{mant95} an original  criterion to perform
multivariate analysis of the stock market historical values was introduced.
The authors were mainly concerned with the problem of deriving a hierarchical structure among the stocks and to this aim they focused on the mutual influences and similarities inside the market.
In the following, we summarize the theoretical background they introduced to develop their criterion.\\
Consider a set of $N$ time series, which are first properly scaled and then modelled as
realizations of stationary stochastic processes, namely $X_i$, $i=1,\ldots,N$.
For each possible couple $(X_i,X_j)$ an estimate of the correlation index $\rho_{ji}$
is computed, along with the quantity
\begin{align}\label{eq:Mantdistance}
	d_{ji}\doteq d(X_i,X_j)\doteq\sqrt{2(1-\rho_{ji})} \,.
\end{align}
In particular, it is worth observing that the function \eqref{eq:Mantdistance} is a metric \cite{gid08}.
In \cite{mant95} each process $X_i$ is interpreted as a node in a graph and $d_{ji}$ as the weight of the arc between $X_i$ and $X_j$.
Then, the Minimum Spanning Tree (MST) is extracted by the related graph, obtaining a simple connected structure based on the strongest similarities emerged from the correlation analysis.
Notably, this procedure has been successfully exploited to provide a topological analysis of time series in financial markets using daily data (see e.g. \cite{mant99,tum07,nrm07}).\\
Recently, the metric (\ref{eq:Mantdistance}) has been interpreted in terms of a modeling procedure \cite{gid08}.
Indeed, let the possible connections between two nodes (processes) be described by the simple scalar constant $\alpha_{ji}$.
Then, modeling $X_j$ by means of $X_i$ produces the error $e_{ji}$ defined as
$e_{ji}=X_j - \alpha_{ji} X_i$
and, thus, by choosing
\begin{align}\label{eq:mant_gain}
	\alpha_{ji}=\frac{E[X_j^2]}{E[X_i^2]} \,,
\end{align}
it is immediate to observe that
\begin{equation}
	\frac{E[e_{ji}^2]}{E[X_j^2]}=2(1-\rho_{ji})=d_{ji}^2 \,.
\end{equation}
Therefore, the weight function results to be the root square of the ratio between the ``powers'' of the modeling error and the modeled process.
Such an interpretation gives the basis for a deeper comprehension of the original procedure.
Indeed, the distance $d_{ji}$ derives from the modeling error $e_{ji}$, that is computed exploiting the constant gain $\alpha_{ji}$ to represent the relationship between the two processes $X_j$ and $X_i$.
Therefore, such a distance is not suitable to describe analogies more complex than proportional laws.
This reasoning is also confirmed by the observation that $d_{ji}$ is the result of a correlation analysis and that $\rho_{ji}$ can just capture the similarities, which occur at the same time instant.
For instance, it can not properly detect analogies in presence of delays or time shifts.
Hence, let us provide an ideal example to stress this important point.\\
Consider a time-discrete, white and stationary stochastic process $X_a$ with unitary power, i.e. such that:
\begin{align}
\label{eq:whiteness}
	E[X_a(t) X_a(t+\tau)] =
	\left\{
		\begin{array}{ll}
			0	\quad \text{if }~\tau\neq 0 \\ 
		    	1	\quad \text{if }~\tau= 0 ~.\\
		\end{array}
	\right.
\end{align}
Further, let $X_b$ be the process obtained from $X_a$ introducing a one time unit delay
$X_b(t)\doteq X_a(t-1)$.
According to \eqref{eq:Mantdistance} and because of the whiteness feature \eqref{eq:whiteness}, it holds that
$d_{ba}=\sqrt{2}$.
Then, even though the behaviour of $X_b$ can be exactly derived from $X_a$, according to \eqref{eq:Mantdistance} the two processes appear as they were not related at all.
Such a simple example is useful to underline that the original procedure
introduced in \cite{mant95} can not be exploited to catch information between samples
at different time instants.
According to the above reasoning, the correlation-based analysis turns out to be more suitable when one deals with data sampled at a low frequency, since in such a situation small delays pass unseen.
Conversely, at high frequencies, we expect a correlation-based analysis not to perform the best.
In \cite{gid08} the authors propose a different multivariate analysis, that mainly extends the correlation approach in order to properly handle the presence of delays, so to detect similarities even at different time instants.\\
The generic time series $X$ can be equivalently represented by its $\mathcal{Z}$-transform, defined as  $\hat X(z)\doteq \sum_{n=-\infty}^{+\infty}X(n)z^{-n}$.
Exploiting the $\mathcal{Z}$-transformation operator, any linear dependence, even
the ones involving different time instants,
can be transformed into an algebraic equation in the variable $z$.
Thus, the modeling error can be defined employing linear dynamical models, instead of considering a simple scalar constant as in \eqref{eq:mant_gain}.
Within this paradigm, a new metric is defined:
\begin{align}\label{eq:distance}
	d(X_i,X_j)\doteq
			\left[ \frac{1}{2\pi}
				 \int_{-\pi}^{\pi}
				\left(1- 
					C_{X_{i} X_{j}}(\w)
				\right) d\w 
			\right]^{1/2} \,,
\end{align}
where 
$C_{X_{i} X_{j}}(\w)=\frac{\Phi_{X_j X_i}^2(\w)}{\Phi_{X_{i}}(\w)\Phi_{X_{j}}(\w)}$
is the well-known coherence function defined in terms of the spectral densities
$\Phi_{X_{i}}(\w)$, $\Phi_{X_{j}}(\w)$ and $\Phi_{X_j X_i}(\w)$.
The distance \eqref{eq:distance} should be apter to describe relations 
among high frequency financial time series, since it is a direct byproduct
of a Wiener identification procedure, i.e. it provides the optimal minimization error
in the sense of the least squares \cite{ksh00}.
Further, it is worth noticing that the coherence function has already been successfully employed in econometrics to perform statistical analysis devoted to quantify the information shared by two time series \cite{gra69}.
\section{Stock market analysis}
\label{sec:exp res}
\begin{figure*}
\begin{tabular}{c}
	\includegraphics[width=0.8\textwidth]{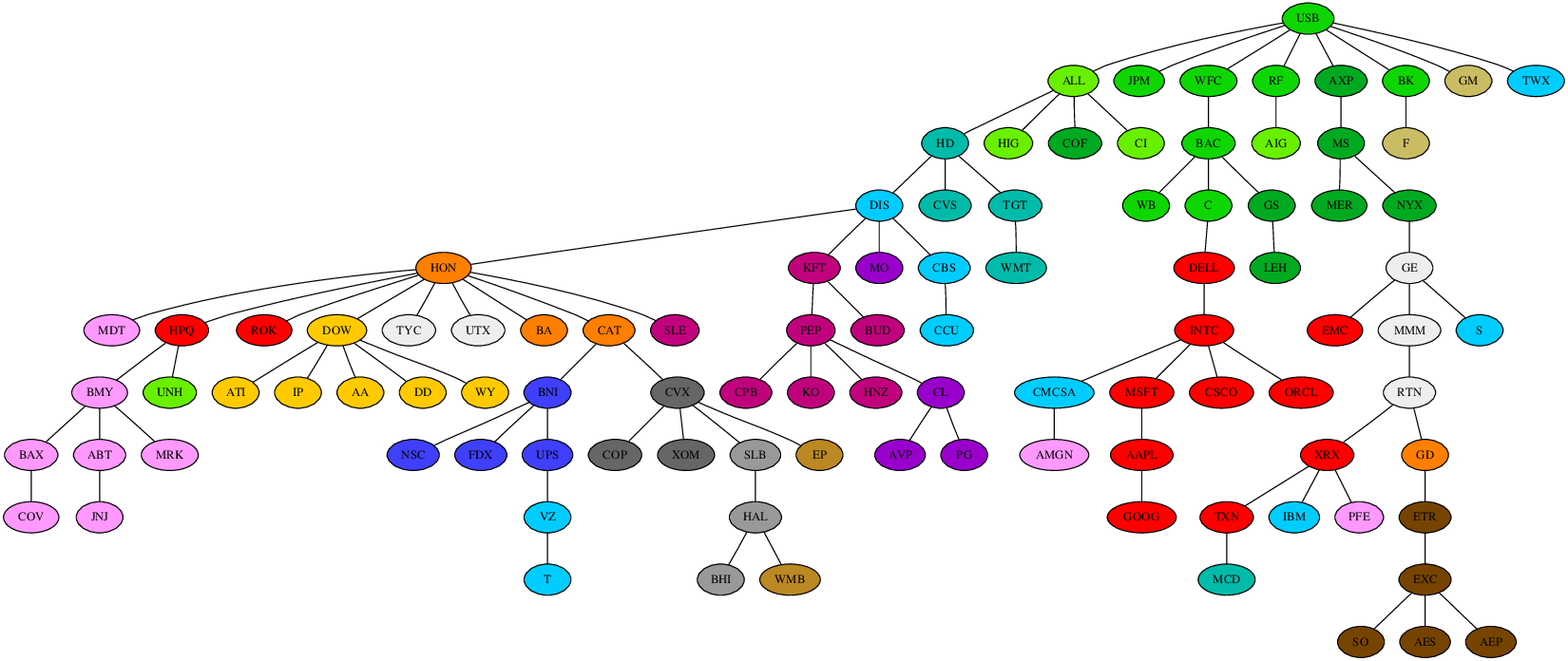}\\ (a)\\
	\includegraphics[width=0.9\textwidth]{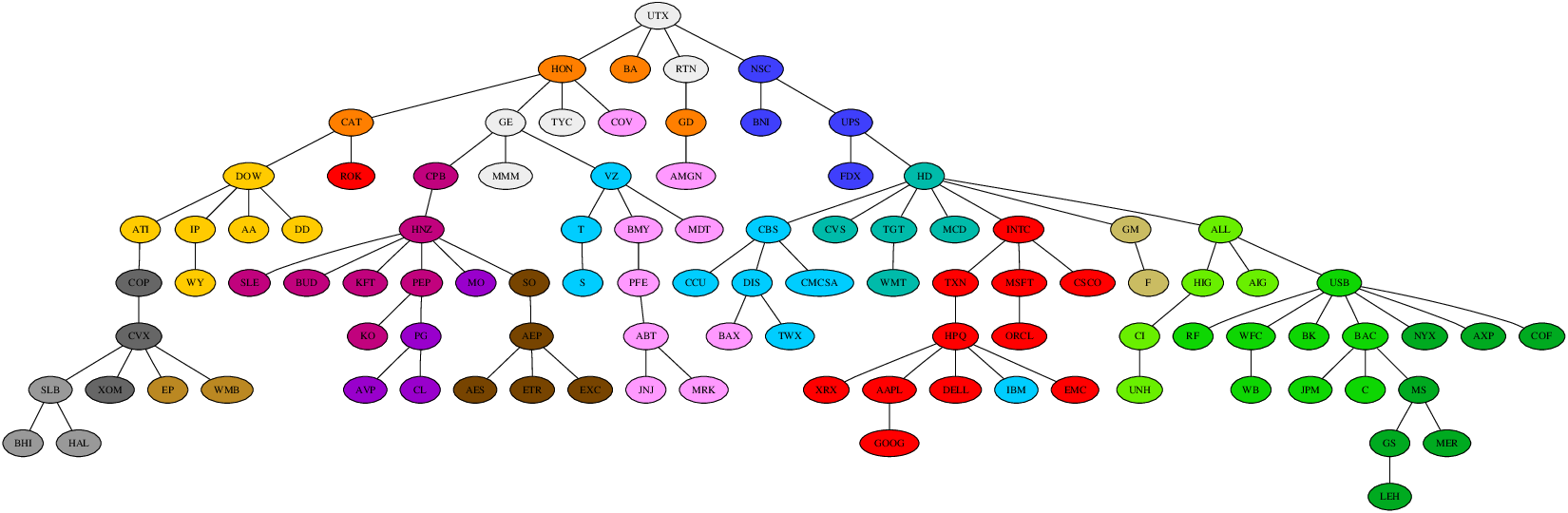}\\ (b)\\
\end{tabular}
	\caption{
		The tree structure obtained using a correlation-based analysis (a) and a coherence based one (b).
		Every node represents a stock and the color represents the business sector it belongs to.
		The considered sectors are
		\texttt{Basic Material} (yellow),
		\texttt{Conglomerates} (white),
		\texttt{Healthcare} (pink),	
		\texttt{Transportations} (dark blue),
		\texttt{Technology} (red),
		\texttt{Capital Goods} (orange),
		\texttt{Utilities} (brown tints),
		\texttt{Consumer} (violet tints),
		\texttt{Financial} (green tints),
		\texttt{Energy} (gray tints)
		\texttt{Services} (light blue tints).
		Using the industry classification given by Google, the \texttt{Financial} sector has also been differentiated among
		Insurance Companies (light green),
		Banks (average green) and
		Investment Companies (dark green);
		\texttt{Services} have been divided in
		Information Technology (cyan) and
		Retail (aquamarina), 
		\texttt{Consumer} in
		Food (plum) and
		Personal-care (purple);
		\texttt{Energy} in
		Oil \& Gas (dark gray) and
		Well Equipment (light gray);
		Utilities in Electrical (dark brown) and Natural Gas (light brown).
		The comparison between (a) and (b) underlines a better capability of the coherence-based method in grouping stocks not only according to their sector, but also according their industry.
		\label{fig:tree_mant_dg} }
\end{figure*}

A collection of $100$ stocks of the New York Stock Exchange have been observed for four weeks (twenty market days),  in the lapse 03/03/2008 - 03/28/2008 sampling their prices every $2$ minutes.
The stocks have been chosen on the first $100$ stocks with highest trading volume according to the Standard \& Poor Index at the first day of observation.
An a-priori organization of the market has been assumed in accordance with the sector and industry group classification provided by Google Finance$^{\text{\textregistered}}$, that is also the source of our data.
We underline that in this paper we are mainly concerned with sectors, but in some cases we have also taken into account the industry group classification to refine the results.
In the following, we introduce a strategy suitable for the application of the method presented in the previous section to the high frequency scenario.\\
The whole observation horizon spans the month of March.
Hence, the corresponding price series can not be considered stationary and the statistical tools can not be successfully employed to analyze the raw data.
In literature a variety of techniques for the suppression of trends and periodic components in non-stationary time series exists.
However, we want to stress that the application of such procedures introduces an additional prefiltering phase, which is responsible for the computational burden increase.
Conversely, hereafter we present a method to avoid data prefiltering,
obtaining a reduction of computations.\\
It is worth observing that the observation horizon is naturally divided into subperiods, namely weeks and days.
Indeed, due to the pre- and post-market sessions, there is a discontinuity between the end value of a day and the opening price of the next one.
Moreover, a single market session can be considered a time period sufficiently short to assume that the influence of trends and seasonal factors are negligible.
Thus, in our analysis, we have followed the natural approach of dividing the historical
series into twenty subperiods corresponding to single days.
Then, we have performed the multivariate analysis for each of these sessions, i.e. we have computed the correlation-based distance \eqref{eq:Mantdistance} and the coherence-based one \eqref{eq:distance} among the stocks.
Finally, we have averaged such distances over the whole observation horizon and the related results have been exploited to extract the MST, providing the corresponding market structure.
We find useful to remark that the computation of the distances for small data sets is best performing and that the averaging procedure provides the desired rejection of trends and seasonal components.
Notably, a similar idea, even if more sophisticated, is at the basis of the method developed in \cite{pos08} to detrend non-stationary time series.
A comparison of the results obtained using the two metrics
is shown in Figure \ref{fig:tree_mant_dg}.
Every node represents a stock and the color represents the business sector or industry it belongs to.
Figure~\ref{fig:tree_mant_dg}a refers to the correlation tree.
We note that the stocks are quite satisfactorily grouped according to their business sectors.
However, as foreseen, we observe an general increased capacity of grouping stocks according to their sectors in the coherence tree of Figure~\ref{fig:tree_mant_dg}b.
It means that the modeling approach of \cite{gid08} has been able to capture time shifted and dynamical dependencies among the time series.
We stress that the a-priori classification in sectors is not a hard fact by itself and we are not trying to match it exactly.
A company could well be categorized in a sector because of its business, but, at the same time, could show a behaviour similar to and explainable through the dynamics of other sectors.
Actually, we would be very interested into finding results of this kind.
Indeed, in those very cases, our quantitative analysis would provide the greatest contributions detecting in an objective way something which is ``counter-intuitive''.
Thus, we just use such a-priori classification as a tool to check if the final topology makes sense and if, at a general level, the coherence approach performs better.
Despite this disclaim, it is worth noting that the \texttt{Financial} (green tints), \texttt{Consumer} (violet tints), \texttt{Basic Materials} (yellow), \texttt{Energy} (gray tints) and \texttt{Transportation} (dark blue) sectors are all perfectly grouped, with no exceptions.
In Figure~\ref{fig:tree_mant_dg}b, we note a subclusterization of the \texttt{Financial} sector, as well.
Such a finer detail can not be detected in the correlation-based tree.
The \texttt{Consumer} sector shows another prominent subclusterization in the \texttt{Food} (plum) and \texttt{Personal/Healthcare} (purple) industries, while
the \texttt{Energy} sector presents an evident subclusterization into the \texttt{Oil \& Gas} (dark gray) and \texttt{Oil Well Equipment} (light gray).
In this case also the correlation approach shows them clearly. In addition both the approaches show the close presence of companies of a different sector and industry, \texttt{Utilities/Natural Gas} (light brown).
The other \texttt{Utilities/Electricity} companies (dark brown) are, interestingly, a different group.
We also observe a big cluster of companies classified as \texttt{Services} (light blue tints).
The correlation tree is not equally capable of grouping these companies together.
We have differentiated them in the two industries \texttt{Retail} and \texttt{Information Technology} using two slightly different colors, respectively aquamarine and cyan.
We also note the presence of three \texttt{Services} companies which are isolated from the other ones: \texttt{V} [Verizon], \texttt{T} [AT\&T], and \texttt{S} [Sprint].
All of them are telephone companies.
This might suggest that this industry should show at least a slightly different dynamics from the other service companies.
Note also how the \texttt{Technology} sector (red) is almost perfectly grouped and how \texttt{IBM}, an IT company, even though classified as a \texttt{Services} company, is located in it.
Finally, the two only automobile companies \texttt{GM} and \texttt{F} [Ford] happen to be linked together.
The analysis of this four weeks of the month of March cleanly shows a taxonomic arrangement of the stocks even though the choice of a tree structure might have seemed quite reductive at first thought.
\section{Conclusion}
We obtain a structural characterization of stocks using high frequency data.
A tree topology is derived, showing a strong taxonomic arrangement of the price time series. Though this property has already been proved for daily prices, to the best of the authors' knowledge, a similar analysis has never been carried on using high frequency data.
The analysis of a collection of $100$ high volume stocks of the New York Stock Exchange has been used to evaluate the results.
A metric based on the coherence function has also been employed to quantify the ``closeness'' of two price historical series.
It is shown to perform consistently better than a standard correlation metric, suggesting the presence of propagative and dynamic phenomena involved in the considered financial network.
Though their presence can not be considered preponderant, they provides information which is not directly captured from a simple correlation analysis.
\bibliography{ifac}
\end{document}